# An optimization problem on the performance of FSO communication system

Mohammad Ali Amirabadi

*Abstract*— Performance of Free Space Optical (FSO) communication system is affected by atmospheric turbulences and pointing errors. These effects can easily be mitigated by adapting natural system parameters such as wavelength. In this paper, considering effects of pointing error and atmospheric turbulence, two optimization models are presented on FSO communication system. In Model 1, the normalized transmitter power is objective function, Bit Error Rate (BER) is equality subjective. In Model 2 normalized transmitter power is equality subjective, the BER is objective function. In both of them the normalized wavelength is variable parameter. These models were previously investigated using numerical methods; in the sense that they were solved asymptotically. From this point of view, this paper regenerated these models and solved them by a completely different analytical method, and derived a new exact solution. Comparing exact and asymptotic methods shows some interesting results; the asymptotic method achieves BER=$10^{-9}$, but the presented new exact method achieves BER=$10^{-13}$, which is a significant difference. It means that even at the worst case scenario such as pointing error, it is possible to have a very good performance by only the transmitting wavelength a bit. Proposed models are practical and obtained results show that this system is cost and power effective.

Authors are with the School of Electrical Engineering, Iran University of Science and Technology (IUST), Tehran 1684613114, Iran A(email: m_amirabadi@elec.iust.ac.ir, vakili@iust.ac.ir)

# I. INTRODUCTION

In FSO system, diodes and lasers are used as transmitters. These light sources emit at infrared bandwidth between 850 and 950 nm. At these frequencies, often Intensity Modulation / Direct Detection (IM / DD) is used. In this scheme the intensity of the transmitted light is modulated. The receiver detects the intensity of the received signal by use of a photo detector and produces an electrical signal proportional to the optical signal intensity. Based on this electrical signal, receiver achieves the original data [1]. Compared with other well-known modulation schemes, On-Off-Keying (OOK) modulation, due to its characteristics is often preferred in IM / DD. In this modulation, detection parameters change according to atmospheric conditions [2, 3].

During the last decades, due to advances and cost reduction in optoelectronic devices, FSO communication systems have attracted many considerations [4]. FSO systems are considered as an appropriate alternative for traditional RF communication systems. They have large bandwidth compared with RF systems. In addition, because of a very narrow beam, FSO systems are highly secure and contains no interference [5]. Also, high data rates demands led to more attention to FSO systems and made them the main competitor of traditional communication systems. RF systems are appropriate in term of cost, but provide lower data rate compared with FSO systems. FSO systems provide both low cost and high data rate [6]. Bandwidth and license constraints of RF systems, which offer data rates up to 100Mbps, encouraged communication companies to find a better solution for the need of bandwidth and data rate. FSO system which provides up to 2000 THz unlicensed bandwidth is an appropriate solution [7].

However, beside FSO advantages, its sensitivity to atmospheric turbulences severely limited its practical implementation [8]. The other main limitation in front of FSO systems is misalignment of transceivers, known as pointing error effect. This effect significantly degrades performance of FSO system, and is divided to zero boresight and non-zero boresight based on vertical and horizontal displacements of transreceiver. In zero boresight, horizontal and vertical displacements on the receiver plane are modeled by zero mean Gaussian distribution, whereas in non-zero boresight, horizontal and vertical displacements are modeled by non-zero mean Gaussian distribution. Radial displacements, in zero boresight error and non-zero boresight error, are modeled by Rayleigh and Rician distributions, respectively. Aperture averaging is a low-cost, simple and useful method to compensate mitigation caused by pointing error [9].

For the transmitter in FSO systems, the effect of building sway is mainly described by the pointing error deviation $\sigma_H$ and $\sigma_V$. So aside system parameters, they need to be included into the FSO performance model. On the other hand, the atmospheric interference such as absorption, scattering, and turbulence involves many more factors. Each type has its own distinct mechanism [10].

Several optimization problems have been done to compensate impairments of FSO systems. A new optimization problem in FSO communication system is developed in [11] considering multi-input multi-output structure. Saturated atmospheric turbulence as well as effect of pointing error are taken into account. Optimization problems on transmitter power and divergence angle at a constant Bit Error Rate (BER) are developed in [12]. However, it did not provide closed-form expressions. Taken into account beam width, pointing error variance, and detector size, an optimization on FSO system in Lognormal and Gamma-Gamma [14] atmospheric turbulences with pointing error is done in [13]. The BER expression for an IM/DD FSO system in strong atmospheric turbulence and pointing error is derived in [15]. [16] considered IM/DD in misalignment model given in [13]. It did not consider atmospheric turbulences.

In [17] two optimization models have developed, and by defining wavelength as variable parameter, these problems were solved. It has investigated the optimization problems in terms of the transmitter power, transmitter wavelength, BER. These optimization problems were solved in [17] using numerical methods; but numerical methods solve asymptotically, therefore, the results do not show the "real" optimal solution. Authors thought it might be interesting to find an exact solution for these problems, and show the difference between these two methods. Therefore, in this paper, the mentioned optimization problems are re-established and re-solved using a new exact mathematical method. Section II is re-establishment of optimization models and is borrowed from [17]; the novelty of this paper is section III, which finds the new exact solution for the mentioned problem. Results in section IV are somehow interesting; it is shown that without power consumption or processing complexity, can get a very good performance.

## II. OPTIMIZATION PROBLEM

As mentioned before, this section is a re-establishment of the proposed problems and is borrowed (not exactly the same) from [17]. This paper considers the effect of pointing error and atmospheric turbulence on the transmitted signal; therefore, considering transmitter power ($P_T$), transmitter and receiver gains ($G_T, G_R$), as well as losses, the received power becomes as follows:

$$P_R = P_T \eta_T \eta_R \left(\frac{\lambda}{4\pi d}\right)^2 G_T G_R L_A L_T, \tag{1}$$

where $\eta_T$ and $\eta_R$ are transmitter and receiver optical efficiencies, $\lambda$ is wavelength, $d$ is transmitter to receiver distance, $L_A$ is atmospheric loss, and $L_T = \exp(-G_T \theta^2)$ is transmitter pointing loss in the case of a Gaussian beam. The term in parentheses is the free-space loss [18], and $G_R \approx (\pi D_R/\lambda)^2$ is receiver telescope gain [19].

At the detector, $P_R$ is converted to electric current as $I_R \approx \rho P_R$, where $\rho$ is the detector's responsivity [11]. In IM/DD with OOK modulation, BER is calculated from:

$$P_e = \frac{1}{2}\int_0^\infty erfc\left(\frac{\rho P_R}{2\sqrt{2}\sigma_N}\right) f(\theta) d\theta, \tag{2}$$

where $\sigma_N^2$ is additive white Gaussian noise variance, and $erfc(.)$ is complementary error function. $P_R$ depends on $\theta$, which is Rayleigh random variable. $f(.)$ is probability density function [17].

Substituting $f(\theta)$ and after some mathematical manipulations, the BER becomes as:

$$P_e = \frac{1}{2}\int_0^\infty erfc(vz\exp(-2xz^2))\exp(-x)dx, \quad (3)$$

where $v = \left(\left(\frac{\rho}{2\sqrt{2}\lambda\sigma_N}\right)\eta_T\eta_R\left(\frac{D_R}{4d}\right)^2 L_A P_T\sqrt{G_T}\right)/\sigma$ is normalized transmitter power, $z = \sigma\sqrt{G_T} = (\pi D_T\sigma)/\lambda$ is normalized transmitter wavelength, $D_T$ and $D_R$ are transmitter and receiver aperture diameters, respectively and $\sigma$ is radial pointing error variance [17].

After some mathematical simplifications, (3) becomes in the following form [17]:

$$P_e = \frac{1}{4}\int_0^1 erfc(vzt^{z^2})t^{-\frac{1}{2}}dt. \quad (4)$$

Differentiating (4), obtains [17]:

$$\frac{dP_e}{dz} = -\frac{v}{2\sqrt{\pi}}\int_0^1 t^{z^2-\frac{1}{2}}(1+2z^2\ln t)\exp(-v^2z^2t^{2z^2})dt. \quad (5)$$

In Model 1, the normalized transmitter power is objective function, BER is equality subjective function. In Model 2 normalized transmitter power is equality subjective, the BER is objective function. In both of them the normalized wavelength is variable parameter. The first optimization problem can be written in the following form [17]:

$$\min v \quad (6)$$
$$s.t. \quad g := P_e - C_1 = 0.$$

According to the definition, $v = f_v(P_e, z)$. Therefore for flinging minimum of $v$ in terms of $z$ should find the root of:

$$\frac{dv}{dz} = -\frac{\frac{dg}{dz}}{\frac{dg}{dv}} = -\frac{\frac{dP_e}{dz}}{\frac{dP_e}{dv}}. \tag{7}$$

Minimization of $v$ in terms of $z$ is equivalent to minimization of $P_e$ in terms of $z$, because $dv/dz = 0$ if and only if $\partial P_e/\partial z = 0$ [17].

The second optimization problem can be written in the following form [17]:

$$\min P_e, \tag{8}$$
$$s.t. \quad v - C_2 = 0.$$

This problem can easily be solved by finding the root of $dP_e/dz = 0$, subject to the constraint. Prior to making effort to solve optimization problem, it has worth to mention that [17] has discussed about existence of solutions which might be interesting to the readers.

## III. EXACT OPTIMIZATION SOLUTION

The solution process of mentioned optimization problems involves evaluating the following equations (4), and (5).

The above integrations can be solved using numerical methods; but this is not the right solution, because numerical methods solve problems asymptotically and cannot find exact optimum solution. In this paper a new exact solution for such integration is presented; this answer will converge to the "real" optimum.

After simplification of (4), and using [20, Eq.06.27.21.0009.01], (4) can be written in the following form:

$$P_e = \frac{1}{2}\int_0^1 erfc(vzu^{2z^2})du = \left[\frac{u}{2}erfc(vzu^{2z^2}) - \frac{vz}{2\sqrt{\pi}}u^{2z^2+1}(v^2z^2u^{4z^2})\Gamma\left(\frac{2z^2+1}{4z^2},v^2z^2\right)\right]_0^1 = \quad (9)$$

$$\frac{1}{2}erfc(vz) - \frac{1}{2\sqrt{\pi}}(vz)^{-\frac{1}{2z^2}}\Gamma\left(\frac{2z^2+1}{4z^2},v^2z^2\right).$$

Differentiating (9), obtains:

$$\frac{dP_e}{dz} = -\frac{v}{\sqrt{\pi}}exp(-(vz)^2) + \frac{1}{2\sqrt{\pi}}\left(-\frac{\ln(vz)}{z^3} + \frac{1}{2z^3}\right)(vz)^{-\frac{1}{2z^2}}\Gamma\left(\frac{2z^2+1}{4z^2},v^2z^2\right) - \quad (10)$$

$$\frac{1}{2\sqrt{\pi}}(vz)^{-\frac{1}{2z^2}}\Gamma_z\left(\frac{2z^2+1}{4z^2},v^2z^2\right),$$

where considering the definition [20, Eq.06.06.02.0001.01], and using [21,Eq.6-40], and [22,Eq.1.6.10] :

$$\Gamma_z\left(\frac{2z^2+1}{4z^2},v^2z^2\right) = -2vz^2(vz)^{\frac{1}{2z^2}}e^{-v^2z^2} - \frac{1}{2z^3}\left[\gamma\left(\frac{2z^2+1}{4z^2},t\right)\ln t - \left(\frac{2z^2+1}{4z^2}\right)^{-2}t^{\frac{2z^2+1}{4z^2}} \times \right.$$

$$\left. {}_2F_2\left(\frac{2z^2+1}{4z^2},\frac{2z^2+1}{4z^2};\frac{2z^2+1}{4z^2}+1,\frac{2z^2+1}{4z^2}+1;-t\right)\right]_{v^2z^2}^{\infty}.$$

(11)

## IV. SIMULATION RESULTS

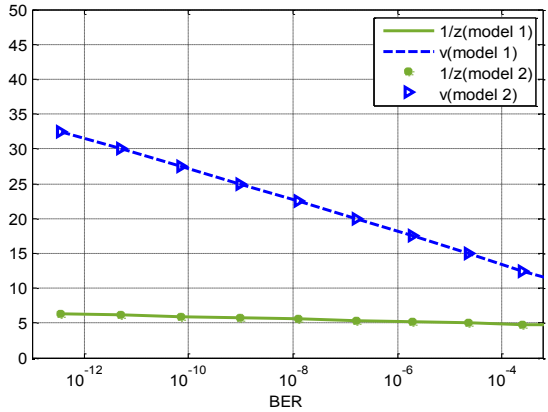

Fig. 1. Exact solution of first and second problems

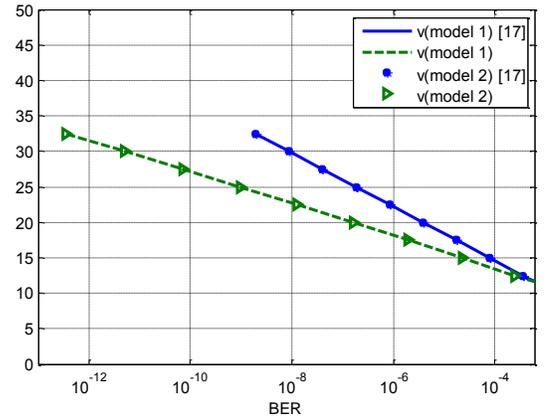

Fig. 2. Exact and asymptotic solution of first and second problems

The minimization problems of (6) and (8) could be solved by finding the roots of $P_e - C_1 = 0$ in (9) and $dP_e/dz = 0$ in (10), respectively. MATLAB Optimization Toolbox, is used to find these roots. The optimal wavelength and the minimum transmitter power are related to the obtained optimum normalized wavelength by: $\lambda_{opt} = (\pi D_T \sigma)/z^*$ and $P_{T,min} = \left(\sigma v^* / \left(\left(\frac{\rho}{2\sqrt{2}\lambda\sigma_N}\right) \eta_T \eta_R \left(\frac{D_R}{4d}\right)^2 L_A \pi D_T\right)\right)$; all of the required parameter values are clarified in [17].

The exact solutions for the first and second problems are presented in Fig. 1. In Model 1, the normalized transmitter power is objective function, BER is equality subjective function, and normalized wavelength is variable parameter. In Model 2 BER is objective function, normalized transmitter power is equality subjective function, and the normalized wavelength is variable parameter. BER values of Model 2 are plotted on the abscissa axis, to compare them with Model 1.

The exact and asymptotic solutions of Model 1 and Model 2 are presented in Fig. 2. As can be seen, e.g. at $v^* = 32.5$, asymptotic solution obtains BER=$10^{-9}$, but exact solution obtains BER=$10^{-13}$. Which shows that the asymptotic solution did not converge to the optimal point, but the exact solution converged.

## V. CONCLUSIONS

In this paper, new exact solutions are presented for two optimization problems developed on performance of FSO system. These models were previously solved asymptotically; but in this paper a new exact solution that completely differs from the previous one is presented. In Model 1, the normalized transmitter power is objective function, BER is equivalent subjective function. In Model 2 normalized transmitter power is equivalent subjective function, the BER is objective function. In both of them normalized wavelength is variable parameter. Obtained exact results are compared with asymptotic results. It is shown that for example at $v^* = 32.5$, asymptotic solution obtains BER=$10^{-9}$, but exact solution obtains BER=$10^{-13}$. In the scene that the asymptotic solution, due to the numerical integration calculations, is unable to find the optimal point; but the exact solution does find it.